\documentclass[9pt]{IEEEtran}

\pdfoutput=1
\usepackage{cite}
\usepackage{amsmath,amssymb,amsfonts}
\usepackage{algorithmic}
\usepackage{graphicx}
\usepackage{textcomp}
\def\BibTeX{{\rm B\kern-.05em{\sc i\kern-.025em b}\kern-.08em
    T\kern-.1667em\lower.7ex\hbox{E}\kern-.125emX}}

\usepackage[hidelinks]{hyperref}
\usepackage{arydshln}
\usepackage{subfigure}
\usepackage{xcolor}
\usepackage{cases}
\newtheorem{assumption}{Assumption}
\newtheorem{remark}{Remark}
\newtheorem{definition}{Definition}
\newtheorem{theorem}{Theorem}
\newtheorem{lemma}{Lemma}

\newtheorem{proposition}{Proposition}

\newcommand{\FF}{\mathbb{F}}
\newcommand{\RR}{\mathbb{R}}
\newcommand{\GG}{\mathbb{G}}

\newcommand{\VV}{\mathbb{V}}

\newcommand{\ones}{\mathbf{1}}
\newcommand{\zeros}{\mathbf{0}}

\newcommand{\bL}{\mathbf{L}}

\newcommand{\cA}{\mathcal{A}}
\newcommand{\cC}{\mathcal{C}}
\newcommand{\cE}{\mathcal{E}}
\newcommand{\cF}{\mathcal{F}}

\newcommand{\cG}{\mathcal{G}}

\newcommand{\cV}{\mathcal{V}}
\newcommand{\cL}{\mathcal{L}}

\newcommand{\cI}{\mathcal{I}}
\newcommand{\cR}{\mathcal{R}}

\newcommand{\cT}{\mathcal{T}}

\begin{document}

\title{Group Consensus of Linear Multi-agent Systems under Nonnegative Directed Graphs}
\author{Zhongchang Liu, ~\IEEEmembership{Member,~IEEE,} Wing Shing Wong, ~\IEEEmembership{Life Fellow,~IEEE}%
\thanks{*This work is supported by the National Natural Science Foundation of China (61703445), Natural Science Foundation of Liaoning Province (20180540064), Innovation Support Program for Dalian High-level Talents (2019RQ057), Dalian Science and Technology Innovation Fund (2019J12GX040), and the Hong Kong Innovation and Technology Fund (ITS/066/17FP) under the HKUST-MIT Research Alliance Consortium.}
\thanks{Z. Liu is with College of Marine Electrical Engineering, Dalian Maritime University, Dalian, 116026, P. R. China (Email: zcliu@foxmail.com).}%
\thanks{W.~S. Wong is with Department of Information Engineering, The Chinese University of Hong Kong, Shatin, N.T., Hong Kong (Email: wswong@ie.cuhk.edu.hk).} %
}
\maketitle

\begin{abstract}
Group consensus implies reaching multiple groups where agents belonging to the same cluster reach state consensus. This paper focuses on linear multi-agent systems under nonnegative directed graphs. A new necessary and sufficient condition for ensuring group consensus is derived, which requires the spanning forest of the underlying directed graph and that of its quotient graph induced with respect to a clustering partition to contain equal minimum number of directed trees. This condition is further shown to be equivalent to containing cluster spanning trees, a commonly used topology for the underlying graph in the literature. Under a designed controller gain, lower bound of the overall coupling strength for achieving group consensus is specified. Moreover, the pattern of the multiple consensus states formed by all clusters is characterized when the overall coupling strength is large enough.
\end{abstract}

\begin{IEEEkeywords}
Group consensus; coupled linear systems; directed spanning trees; graph topology
\end{IEEEkeywords}

\section{Introduction}
Multi-agent systems (MASs) have been continuingly attracting research attentions and have found wide applications \cite{Chen&Ren2019}. While prevalent works concentrate on reaching global consensus/synchronization for all agents, there arises increasing interest in the problem of group consensus (or cluster consensus, group/cluster synchronization), where coupled systems converge to multiple synchronous groups instead of one. This research topic is mainly motivated from multi-modal opinion dynamics in social networks \cite{Li&Scaglione2013SocialNetworks} and clustering of oscillatory networks \cite{Qin2004CS}, and have potential applications in power grids and multiple interconnected formations \cite{Gexiaohua2018}.

For the problem of reaching global consensus, comprehensive understandings about the underlying graph topologies and control algorithms subject to various agent dynamics have been established \cite{Ma&JFZhang2010,Li&GRChen2010}. In contrast, the mechanisms for achieving group consensus have not been fully understood yet. Early works such as \cite{Wu&Chen2009,Yu_Wang2010,Xia2011} presented sufficient algebraic conditions on the graph Laplacian for achieving a prescribed cluster consensus pattern by assuming balanced positive and negative weights for inter-cluster links. Subsequent works under this framework mainly designed distributed control algorithms for different types of agent dynamics \cite{Qin&Yu2013,Yu&Qin2014,Qin2016,Liu&Wong_clustering2017,Qin2017TIC,Qin2018} and provided lower bounds for  intra-cluster coupling strengths, in which the underlying topology of each cluster is required to contain a spanning tree.
On the other hand, for MASs that have all edge weights being positive, the in-degrees of all nodes in the same cluster from any other cluster should be equal (i.e., the so-called \emph{inter-cluster common influence condition}) so as to maintain the group consensus manifolds invariant \cite{Lu&Chen2010}. Under this framework, the underlying topology that contains \emph{cluster spanning trees} was proved to be necessary and sufficient under undirected graphs for chaotic oscillators in \cite{Lu&Chen2010}, and under balanced digraphs for discrete-time single integrators in \cite{Shang2016}. For general nonnegatively weighted digraphs, this topology was taken as a sufficient condition when enforcing cluster consensus for coupled single integrators \cite{Han&Chen2013,Han&Chen2015}. It is also necessary when the structure of inter-cluster connections does not form any cycle  \cite{Liu&Wong_CCC2016,Liu&Wong_CDC2016}. However, the necessity for general nonnegative digraphs remains unconfirmed to the best knowledge of the authors. 
Other relevant works considered the group consensus patterns that may emerge in undirected networks or unweighted digraphs from perspectives including group theory \cite{Pecora2014Nature,Sorrentino&Pecora2016,Klickstein&Pecora2019} and graph partitions \cite{Monaco2019} without specifying connectivities of the underlying graph topology. In summary, although remarkable results have been reported, there still lacks a unified knowledge about the necessary features of underlying digraphs for ensuring group consensus.
In addition, the characteristics of the consensus states in the clusters are rarely specified except for MASs with simple individual dynamics \cite{Monaco2019}.

In undirected and unweighted networks, the quotient graph associated with an external equitable partition (which is equivalent to the inter-cluster common influence condition) of the underlying full graph is shown to govern motions of the group consensus manifold \cite{OClery2013,Schaub2016}.
In \cite{Gambuzza&Frasca2019,Gambuzza&Frasca2020}, sufficient conditions in terms of eigenvalues of the quotient graph Laplacian and the full graph Laplacian are presented to ensure stabilities of group consensus manifolds.
This paper will generalize the definition of quotient graph to weighted digraphs, and show how its connectivity and its Laplacian can actually determine group consensusability.
Unlike the quotient graph of a connected undirected graph which is inherently strongly connected \cite{OClery2013}, the quotient graph of a weakly connected digraph can contain multiple branches, resulting in multiple zero eigenvalues in its Laplacian. Hence, graph theories established in \cite{Caughman&Veerman2006} and \cite{Wu2007} for such topologies will be introduced to lay partial foundations of this work.

Focusing on reaching group consensus for generic linear MASs under nonnegatively weighted digraphs, this paper presents the following main results: a)  A new necessary and sufficient graph topology is derived, which requires the spanning forest of the underlying digraph and that of its quotient graph to contain equal minimum number of directed trees.  This condition is verifiable without examining the connection details inside any cluster compared with the existing condition of containing cluster spanning trees. b) The Laplacian of the quotient graph is shown to be decomposable from the full graph Laplacian, while the eigenvalues of the remaining part determines group consensusability and specifies the overall coupling strength for ensuring group consensus under a separately designed controller gain. This result strengthens and generalizes partial conclusions for MASs under undirected or unweigted graphs in \cite{Gambuzza&Frasca2019,Gambuzza&Frasca2020}. c) The final group consensus states are characterized, which will be consistent with the pattern of MASs with single integrator dynamics in \cite{Monaco2019} if the overall coupling strength is large enough, while may not otherwise.

\emph{Notation}:
$\ones_{n}=[1,1,\ldots,1]^T\in \RR^{n}$. $I_n$ is the identity matrix of dimension $n$. $blkdg\{M_1,\ldots, M_n\}$ represents the block diagonal matrix constructed by matrices $M_1,\ldots,M_n$.
The symbol ``$\otimes$" stands for the Kronecker product.
For a square matrix $M$, its spectrum is denoted by $\sigma(M)$, and the real part of its eigenvalue is denoted by $Re\lambda(M)$.
\section{Problem Statement}\label{sec:Problemstatement}
Consider a multi-agent system (MAS) consisting of $L$ agents indexed by the set $\cI=\{1,\ldots,L\}$.
The individual dynamics of each agent is described by the following generic linear system model
\begin{equation}
\label{sys:linearmodel}
\dot{x}_l(t) =Ax_l+Bu_l(t),\; l\in \cI,
\end{equation}
where $x_l(t)\in \RR^n$ is the state of agent $l$ with initial value $x_l(0)$, $u_l(t)\in \RR^{n_u}$ is the control input, $A\in\RR^{n\times n}$, $B\in\RR^{n\times n_u}$, and $(A,B)$ is a stabilizable pair.%

These $L$ agents belong to $N$ distinct clusters denoted by the index sets $\cC_i$, $i=1,\ldots,N$.
Assume without loss of generality that each cluster $\cC_i$ contains $l_i\geq 1$ agents ($\sum_{i=1}^{N}l_i=L$), and the indices are arranged such that $\cC_1=\{1,\ldots, l_1\},\ldots$, $\cC_i=\{\rho_i+1,\ldots, \rho_i+l_i\}, \ldots$, $\cC_N=\{\rho_N+1,\ldots, \rho_N+l_N\}$, where $\rho_1=0$ and $\rho_i=\sum_{j=1}^{i-1}l_j, \;i=2,\ldots, N$.
Hence, the set $\cC=\{\cC_1,\ldots,\cC_N\}$ is a nontrivial partition of the index set $\cI$, and is called a \emph{clustering} of the above multi-agent system. 
Two distinct agents, $l$ and $k$ in $\cI$, are said to belong to the same cluster $\cC_i$ if $l\in\cC_i$ and $k\in\cC_i$.

\subsection{The Group Consensus Problem}

In this study, the agents are supposed to be linearly coupled through their control inputs:
\begin{align}\label{sys:controllaws_static}
u_l(t)=\delta K\sum_{k\in\cI}w_{lk}\left[x_k(t)-x_l(t)\right],l\in\cC_i, i=1,\ldots,N
\end{align}
where $\delta>0$ is the overall coupling strength used to compensate for the underlying topology, $K$ is the controller gain matrix to be determined, and $w_{lk}\ge 0$ is the weight of the link from agent $k$ to agent $l$. 
Then the closed-loop equations for \eqref{sys:linearmodel} are described by
\begin{equation} \label{eq:MAS_closedloop}
\dot{x}_l(t) = Ax_l-\delta BK\sum_{k=1}^{L}\ell_{lk}x_k(t),\; l\in \cC_i,\; i=1,\ldots,N
\end{equation}
where $\ell_{ll}=\sum_{k\in\cI} w_{lk}$ and $\ell_{lk}=-w_{lk}$ for any $k\neq l$.
Concatenating variables in $x(t)=[x^T_1(t),\ldots,x^T_{L}(t)]^T\in\RR^{nL}$, 
we can write \eqref{eq:MAS_closedloop} into the following compact form:
\begin{equation} \label{eq:MAS_compact}
\dot{x}(t) = (I_L\otimes A-\delta\cL\otimes BK) x(t),\; x(0)\in\RR^{nL}
\end{equation}
where $\cL=[\ell_{lk}]$. 

The group consensus problem is defined below \cite{Yu_Wang2010,Qin2017TIC}.
\begin{definition} \label{def:cluster_syn}
The  multi-agent system in \eqref{eq:MAS_compact} achieves \emph{group consensus} with respect to (w.r.t.) the clustering $\cC$ if for any $x_l(0)\in\RR^n$, $l\in\cI$,
$\lim_{t\rightarrow\infty}\|x_{l}(t)-x_{k}(t)\|=0$, $\forall k,l \in\cC_i$, $i=1,\ldots,N$. 
\end{definition}

Note that group consensus do not require the consensus states in different clusters to be distinct \cite{Yu_Wang2010,Qin2017TIC}, which are equivalent to the definition of intra-cluster consensus in cluster consensus problems \cite{Han&Chen2013,Han&Chen2015,Lu&Chen2010}. Considering that group consensus is the prerequisite of reaching cluster consensus while state separations for different clusters can be enforced by some extra techniques as in \cite{Han&Chen2013,Han&Chen2015}, our work will focus on the fundamental problem of group consensus only.

It is trivial to see that the group consensus problem can be solved if the MAS can achieve global consensus for their states, i.e., $\lim_{t\rightarrow\infty}\|x_{l}(t)-x_{k}(t)\|=0$, $\forall k,l \in\cI$. However, global consensus is only a special case of group consensus. The goals of this paper are to reveal general graph topologies that can ensure group consensus for the MAS \eqref{eq:MAS_compact}, and to further shed some light on the patterns of the achieved multiple consensus states. 

\subsection{Useful Graph Theory}
A directed graph (digraph) $\cG=(\cV,\cE,\cA)$ is associated with the MAS (\ref{sys:linearmodel}) such that each agent is considered as a node in the node set $\cV$, while connections among agents correspond to directed edges in $\cE\subset\cV\times\cV$.
The adjacency matrix $\cA=[w_{lk}]$ is defined such that $w_{lk}> 0$ if there is a directed edge from agent $k$ to agent $l$, and $w_{lk}=0$, otherwise. 
The in-degree of a node $l$ is the quantity $\sum_{k\in\cI}w_{lk}$. The Laplacian matrix of $\cG$
is $\cL=[\ell_{lk}]$ with each entry $\ell_{lk}$ being defined in \eqref{eq:MAS_closedloop}. The digraph $\cG$ is weakly connected if the graph derived via replacing all directed edges of $\cG$ with undirected edges is connected. A \emph{directed spanning tree} of $\cG$ is a directed tree that contains all the nodes through directed paths in $\cG$. A \emph{directed spanning forest} of $\cG$ is a digraph consisting of one or more directed trees that together contain all the nodes of $\cG$, but no two of which have a node in common.
$\cG$ is said to contain \emph{cluster spanning trees} w.r.t. the clustering $\cC$ if for each cluster $\cC_i$, $i=1,\ldots,N$, there exists a node in $\cV$ which can reach all nodes with indices in $\cC_i$ through directed paths in $\cG$. Note that the paths used to span a cluster of nodes may contain nodes belonging to other clusters, and of course can also follow inter-cluster edges.


It is well-known that a strongly connected graph has an irreducible Laplacian matrix.
For graph topologies that are not strongly connected, its Laplacian is reducible. Hence, the following conclusions about an $m$-reducible Laplacian matrix from \cite{Wu2007} will be useful.
\begin{lemma}[\cite{Wu2007}] \label{thm:m_reducible}
Let $M\in\RR^{N\times N}$ be a reducible Laplacian matrix of a nonnegative digraph. The following statements are equivalent for any $m\in\{1,\ldots,N\}$:
 \begin{itemize}
   \item[(a)] $M$ is $m$-reducible.
   \item[(b)] The zero eigenvalue of $M$ has multiplicity $m$, and all the other eigenvalues have positive real parts.
   \item[(c)] $m$ is the minimum number of directed trees which together span the digraph.
 \end{itemize}
\end{lemma}

By this lemma, a $1$-reducible Laplacian corresponds to a graph that is not strongly connected but contains a directed spanning tree.

\begin{figure}[t!]
  \centering
  \subfigure[A graph $\cG$ partitioned into three subgraphs $\cG_1$, $\cG_2$, $\cG_3$.]{
  \includegraphics[width=4cm]{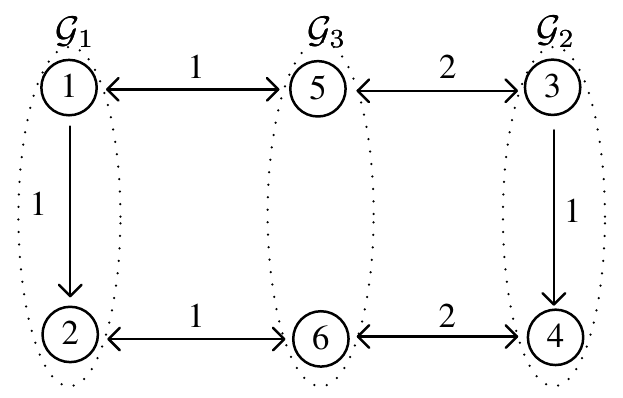}
  \label{fig:topology_acyclic}
  } \hspace{1mm}
  \subfigure[The quotient graph $\GG$ induced from $\cG$ w.r.t. the partition in (a).]{
  \includegraphics[width=4cm]{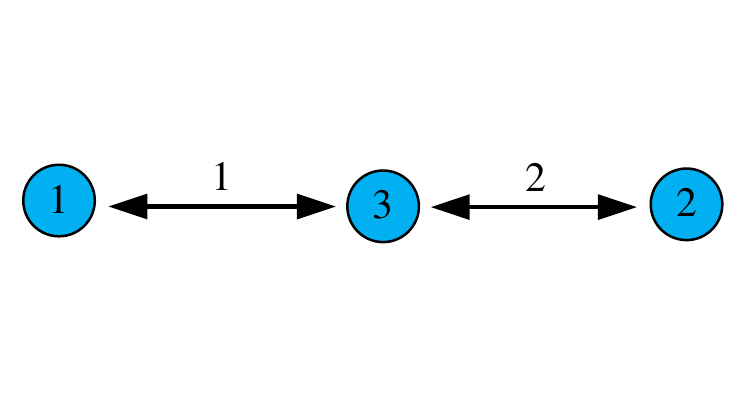}
  \label{fig:topology_induced}
  }
 \caption{Interaction graph and its quotient graph.}
 \label{fig:topology_example1}
\end{figure}

\subsection{Assumptions}
Corresponding to the clustering $\cC=\{\cC_1,\ldots,\cC_N\}$,  let the subgraph of $\cG$, denoted by $\cG_i$, contain all the nodes with indices in $\cC_i$ and the edges connecting them directly (all inter-cluster links are excluded from $\cG_i$, and see Fig.  \ref{fig:topology_acyclic} for an illustration). Then, the Laplacian matrix $\cL$ of $\cG$ can be partitioned into the following block-matrix form:
\begin{equation}
\label{var:laplacianpartition}
\cL=\begin{bmatrix}
L_{11}&L_{12}&\cdots&L_{1N}\\
L_{21}&L_{22}&\cdots&L_{2N}\\
\vdots&\vdots&\ddots&\vdots& \\
L_{N1}&L_{N2}&\cdots&L_{NN}\\
\end{bmatrix},
\end{equation}
where each diagonal block $L_{ii}\in \RR^{l_i\times l_i}$ specifies \emph{intra-cluster} interactions, and each off-diagonal block $L_{ij}\in \RR^{l_i\times l_j}$ with $i\neq j$, $i,j=1,\ldots,N$ specifies \emph{inter-cluster} interactions from nodes in cluster $\cC_j$ to nodes in $\cC_i$.

To ensure group consensus, the Laplacian $\cL$ is assumed to satisfy the following condition.
\begin{assumption} \label{assump:constant_row_sums}
Every block $L_{ij}$ of the Laplacian  $\cL$ defined in \eqref{var:laplacianpartition} has a constant row sum $\beta_{ij}$, i.e. $L_{ij}\ones_{l_j}= \beta_{ij}\ones_{l_j}$, for $i,j=1,\ldots,N$.
\end{assumption}

The above assumption implies that $\sum_{k\in\cC_j}w_{lk}=\sum_{k\in\cC_j}w_{l'k}$ for any $l\neq l'$ in $\cC_i$, and $i\neq j$, i.e.,  the in-degrees of all nodes in a cluster with respect to another cluster are equivalent. Any clustering $\cC$ that renders the graph Laplacian $\cL$ satisfying Assumption \ref{assump:constant_row_sums} is also called an external equitable partition (EEP) of $\cG$ \cite{OClery2013,Schaub2016}.
It has been shown in the literature (such as \cite{Lu&Chen2010,Han&Chen2013,Schaub2016}) that Assumption \ref{assump:constant_row_sums} is necessary for the \emph{group consensus manifold} $\{x(t)\in\RR^{nL}: x_l(t)=x_{k}(t), \forall l,k\in\cC_i,i=1\ldots,N\}$ to be invariant. 
An intuitive reasoning is that under this condition different agents in the same cluster will receive equivalent influence from another cluster in the group consensus manifold. Hence, Assumption \ref{assump:constant_row_sums} is also called the \emph{inter-cluster common influence} condition in \cite{Lu&Chen2010,Han&Chen2013,Han&Chen2015}.

In the following presentation, two more basic assumptions are made to exclude trivial cases. One is that $\cG$ is at least weakly connected (i.e. contains no isolated component) so as to exclude the apparently infeasible graph topologies where agents belonging to the same cluster happen to reside in different isolated components of the network. The other assumption is that the system matrix $A$ has at least one eigenvalue located in the closed right half-plane so as to avoid reaching trivial global consensus all the time.

\section{Achieving Group Consensus} \label{sec:syn_leaderless}
This section will establish the conditions for ensuring group consensus for MAS \eqref{eq:MAS_compact} by bridging the connectivity of $\cG$, which describes  inter-agent connections, with the connectivity of its induced quotient graph $\GG$ w.r.t. $\cC$, which describes inter-cluster interactions.
The nomenclature of quotient graph follows from \cite{OClery2013,Schaub2016} for EEPs of unweighted graphs. In the following, we give an intuitive definition for weighted graphs through construction.
\begin{definition} \label{def:induced_graph}
 Given a graph $\cG$ and its partition $\{\cG_1,\ldots,\cG_N\}$ w.r.t. the clustering $\cC$, \emph{the quotient graph $\GG$ induced from $\cG$} is constructed through the following steps:
\begin{enumerate}
  \item collapsing each subgraph $\cG_i$ into a single node with index $i$;
  \item defining a directed edge from node $i$ to node $j$ in $\GG$ if and only if there exists at least one directed edge in $\cG$ pointing from a node in $\cG_i$ to a node in $\cG_j$;
  \item defining the edge weight from node $j$ to node $i$ in $\GG$ as
       \begin{equation}
       \alpha_{ij}=\frac{1}{l_i}\sum_{l\in\cC_i}\sum_{k\in\cC_j}{w_{lk}}, i,j=1,2,\ldots,N.
       \end{equation}
\end{enumerate}
\end{definition}
Under Assumption \ref{assump:constant_row_sums}, each edge weight of $\GG$ will reduce to $\alpha_{ij}=\sum_{k\in\cC_j}{w_{lk}}$ for any $l\in\cC_i$ (see Fig. \ref{fig:topology_induced}). Also, each constant row sum $\beta_{ij}$ defined in Assumption \ref{assump:constant_row_sums} can be computed by $\beta_{ij}=-\alpha_{ij}$ for $i\neq j$, and $\beta_{ii}=\sum_{j=1}^{N}\alpha_{ij}$. It follows that under Assumption \ref{assump:constant_row_sums} the Laplacian of the quotient graph $\GG$ can be defined as follows:
\begin{equation} \label{var:laplacian_induced}
\cL_{\GG}=[\beta_{ij}]_{i,j=1,\ldots,N}.
\end{equation}


\subsection{Necessary And Sufficient Graph Topologies}
For each $i\in\{1,\ldots,N\}$, define $e_l(t)=x_l(t)-x_{\rho_i+1}(t)$ as the state difference between the agent $\rho_i+1$ in cluster $\cC_i$ and any other agent $l\in\cC_i\setminus\{\rho_i+1\}$. It follows from \eqref{eq:MAS_closedloop} that
\begin{align} \label{eq:CS_11}
\dot e_l(t)&= Ae_l-\delta BK\sum_{k=1}^{L}(\ell_{lk}-\ell_{\rho_i+1,k})x_k(t),\notag\\
&=Ae_l-\delta BK\sum_{j=1}^{N}\sum_{k\in\cC_j}(\ell_{lk}-\ell_{\rho_i+1,k})[x_k(t)-x_{\rho_j+1}(t)]\notag\\
&=Ae_l-\delta BK\sum_{j=1}^{N}\sum_{k\in\cC_j}(\ell_{lk}-\ell_{\rho_i+1,k})e_k(t)
\end{align}
where the second equality is valid since for any $j\in\{1,\ldots,N\}$ and any $x_{\rho_j+1}\in\RR^n$, $\sum_{k\in\cC_j}(\ell_{lk}-\ell_{\rho_i+1,k})x_{\rho_j+1}=(\sum_{k\in\cC_j}\ell_{lk}-\sum_{k\in\cC_j}\ell_{\rho_i+1,k})x_{\rho_j+1}=0$ due to  Assumption \ref{assump:constant_row_sums}.
Stacking the state difference vectors $e_l(t)$, $l\in\cC_i\setminus\{\rho_i+1\}$, $i=1,\ldots,N$ in $$e(t)=[e^T_{\rho_1+2}(t),\dots,e^T_{\rho_1+l_1}(t),
\cdots,e^T_{\rho_N+2}(t),\ldots,e^T_{\rho_N+l_N}(t)]^T,$$ 
one can get from \eqref{eq:CS_11} that
\begin{equation} \label{eq:CS_12}
\dot e(t) =(I_{L-N}\otimes A -\delta\hat\cL\otimes BK) e(t),
\end{equation}
where $\hat\cL\in\RR^{(L-N)\times(L-N)}$ is in the following block-matrix form
\begin{equation}\label{eq:CS_19}
\hat \cL=[\hat L_{ij}]_{i,j=1,\ldots,N},
\end{equation}
with each block $\hat L_{ij}\in\RR^{(l_i-1)\times (l_j-1)}$ being defined by \begin{equation}\label{eq:CS_18c}
\hat L_{ij}=\tilde L_{ij}-\ones_{l_i-1}\gamma_{ij}^T,\;i,j=1,\ldots,N,
\end{equation}
where \begin{align} \label{eq:CS_18a}
\gamma_{ij}&=[\ell_{\rho_i+1,\rho_j+2},\cdots,\ell_{\rho_i+1,\rho_j+l_j}]^T\in\RR^{l_j-1},\\
\label{eq:CS_18b}
\tilde L_{ij}&=\begin{bmatrix}
\ell_{\rho_i+2,\rho_j+2}&\cdots&\ell_{\rho_i+2,\rho_j+l_j}\\
\vdots&\ddots&\vdots \\
\ell_{\rho_i+l_i,\rho_j+2}&\cdots&\ell_{\rho_i+l_i,\rho_j+l_j}\\
\end{bmatrix}\in\RR^{(l_i-1)\times (l_j-1)}.
\end{align}
It is clear from \eqref{eq:CS_12} that group consensus can be achieved for any initial state $x(0)\in\RR^{nL}$ if and only if $I_{L-N}\otimes A -\delta\hat\cL\otimes BK$ is Hurwitz, i.e., the state motions transversal to the group consensus manifold are stable. By using properties of Kronecker products, the stability of $I_{L-N}\otimes A -\delta\hat\cL\otimes BK$ can be ensured by the stabilities of $A-\delta\lambda_{l}(\hat \cL)BK$ for all $\lambda_{l}(\hat \cL)\in\sigma(\hat \cL)$. It follows that $\hat\cL$ plays a key role in rendering group consensus. In the following lemma, we will show that $\hat\cL$ can be decomposed from the full graph Laplacian $\cL$, and its stability can be specified in terms of graph topologies.

\begin{lemma}\label{thm:Laplacian_reduced}
Under Assumption \ref{assump:constant_row_sums}, all eigenvalues of $\hat \cL$ have positive real parts if and only if the spanning forest of $\cG$ and that of the quotient graph $\GG$ contains equal minimum number of directed trees.
\end{lemma}
\begin{IEEEproof}
 Defining $\eta_{ij}=[\ell_{\rho_i+2,\rho_j+1},\cdots,\ell_{\rho_i+l_i,\rho_j+1}]^T$, and using \eqref{eq:CS_18a} and \eqref{eq:CS_18b}, we can represent each $L_{ij}=[\ell_{kl}]_{k=\rho_i+1,\ldots,\rho_i+l_i,l=\rho_j+1,\ldots,\rho_j+l_j}$ in the following block form
$L_{ij}=\begin{bmatrix}
 \ell_{\rho_i+1,\rho_j+1}&\gamma_{ij}^T\\
 \eta_{ij}&\tilde L_{ij}
 \end{bmatrix}$.
Define $S_i=\begin{bmatrix}1 &0\\\ones_{l_i-1} &I_{l_i-1}\end{bmatrix}\in\RR^{l_i\times l_i}$ with inverse $S_i^{-1}=\begin{bmatrix} 1 &0\\ -\ones_{l_i-1} &I_{l_i-1}\end{bmatrix}$ for $i=1,\ldots,N$. Further invoking  $L_{ij}\ones_{l_j}= \beta_{ij}\ones_{l_j}$ from Assumption \ref{assump:constant_row_sums}, one can get that $S_i^{-1}L_{ij}S_j=
\begin{bmatrix}
 \beta_{ij}&\gamma_{ij}\\
 0&\hat L_{ij}
 \end{bmatrix}$.
Let $S=blkdg\{S_1,\ldots,S_N\}$.
Then, one has the following nonsingular transformation
\begin{equation} \label{eq:CS_6}
S^{-1}\cL S=
\begin{bmatrix}
\beta_{11}&\gamma_{11}&\cdots&\beta_{1N}&\gamma_{1N}\\
0&\hat L_{11}&\cdots&0&\hat L_{1N}\\
\vdots&\vdots&\ddots&\vdots&\vdots \\
\beta_{N1}&\gamma_{N1}&\cdots&\beta_{NN}&\gamma_{NN}\\
0&\hat L_{N1}&\cdots&0&\hat L_{NN}
\end{bmatrix}.
\end{equation}
By permutating the columns and rows of $S^{-1}\cL S$ in \eqref{eq:CS_6}, one can get the following block upper-triangular matrix
\begin{equation} \label{var:laplacianpartition_transformed}
\begin{bmatrix}
\cL_{\GG}&[\gamma_{ij}]_{i,j=1,\ldots,N}\\
0_{(L-N)\times N}&\hat \cL\\
\end{bmatrix}
\end{equation}
where $\cL_{\GG}$ is the Laplacian of the quotient graph as defined in \eqref{var:laplacian_induced}, and $\hat \cL$ is defined in \eqref{eq:CS_19}.
It follows that the matrix $\hat \cL$ is nonsingular with all eigenvalues having positive real parts if and only if the two Laplacians $\cL$ and $\cL_{\GG}$ have equal number of zero eigenvalues. Further using Lemma \ref{thm:m_reducible} (b) and (c) yields the conclusion of this lemma.
\end{IEEEproof}


As a stabilizable pair $(A,B)$, for any $Q>0$ there is a positive definite $P>0$ satisfying the following algebraic Riccati equation
\begin{equation} \label{eq:CS_8}
PA+A^TP- PBB^TP=-Q.\;\;
\end{equation}
Using the above, we can derive the main result in the following.
\begin{theorem}\label{thm:intracluster_consensus}
Under Assumption \ref{assump:constant_row_sums}, the MAS \eqref{eq:MAS_compact} can achieve group consensus w.r.t. $\cC$ for any initial state $x(0)\in\RR^{nL}$ if and only if $\cL_{\GG}$ and $\cL$ have equal number of zero eigenvalues, or equivalently, the spanning forest of $\cG$ and that of its quotient graph $\GG$ contain equal minimum number of directed trees.
\end{theorem}
\begin{IEEEproof}
For the sufficiency part, if the conditions in Theorem \ref{thm:intracluster_consensus} hold, then $Re\lambda_{l}(\hat \cL)>0$ for each $\lambda_{l}(\hat \cL)\in\sigma(\hat \cL)$ by Lemma \ref{thm:Laplacian_reduced}. Then one can choose
\begin{equation} \label{eq:CS_14a}
\delta\ge 1/\min_{l} 2 Re\lambda_{l}(\hat \cL),
\end{equation} and let $K=B^TP$ where $P$ is defined in \eqref{eq:CS_8}, such that for each $\lambda_{l}(\hat \cL)\in\sigma(\hat \cL)$, the following Lyapunov inequality holds
\begin{align} \label{eq:CS_7}
&\quad(A-\delta\lambda_{l}(\hat \cL) BK)^*P+P(A-\delta\lambda_{l}(\hat \cL) BK)\notag\\
&=A^TP+PA-2\delta Re \lambda_{l}(\hat \cL) PBB^TP \notag\\
&=-Q-(2\delta Re \lambda_{l}(\hat \cL)-1)PBB^TP\leq-Q.
\end{align}
Hence, $A-\delta\lambda_{l}(\hat \cL)BK$ is Hurwitz for each $l=1,\ldots,L-N$ which implies $I_{L-N}\otimes A -\delta\hat\cL\otimes BK$ is Hurwitz.

On the other hand, 
the violation of the condition in Theorem \ref{thm:intracluster_consensus} implies by Lemma \ref{thm:Laplacian_reduced} that the matrix $\hat \cL$ has at least one zero eigenvalue, i.e., there exists at least one $l^*\in\{1,\ldots,L-N\}$ such that $\lambda_{l^*}(\hat \cL)=0$. It turns out that $A-\delta\lambda_{l^*}(\hat \cL)BK=A$ is not Hurwitz, which implies that group consensus cannot be guaranteed for all initial states.
\end{IEEEproof}

\begin{remark}
The combination of Theorem \ref{thm:intracluster_consensus} and Lemma \ref{thm:Laplacian_reduced} reveals that the positivity of the quantity $\min_{l} Re\lambda_{l}(\hat \cL)$ determines the feasibility of a graph topology for ensuring group consensus, while its value determines the convergence rate of group consensus. The role of this quantity is comparable with that of the minimum real part of nonzero eigenvalues of the Laplacian $\cL$ (i.e., $\min_{l\neq 1} Re\lambda_{l}(\cL)$ where $\lambda_{1}(\cL)=0$) in global consensus problems \cite{Ma&JFZhang2010,Li&GRChen2010}.
\end{remark}
\begin{remark} As noticed in a recent paper \cite{Gambuzza&Frasca2020}, the matrix $\hat \cL$ can be decomposed from $\cL$ by using the characteristic matrix that describes localization of each node in each cluster when the edges are unweighted, while the minimum eigenvalue of $\hat \cL$ is used directly in designing controller gains for MASs with second-order linear dynamics. In comparison, the decomposition method in Lemma \ref{thm:Laplacian_reduced} applies to weighted digraphs, and our approach of specifying the controller gains and the coupling strength separately can deal with generic linear system dynamics.
\end{remark}
\begin{remark}
We also remark that although the conclusion in Theorem \ref{thm:intracluster_consensus} is drawn for linear MASs with identical individual dynamics, extensions to MASs with heterogeneous individual (linear or nonlinear) dynamics are immediate providing that proper controllers can be designed and a large enough coupling strength is chosen by using existing techniques such as those presented in \cite{Liu&Wong_CDC2016,Liu&Wong_clustering2017,Qin2018,Qin2017TIC}.
\end{remark}

\subsection{An Alternative Condition}
The conditions presented in Theorem \ref{thm:intracluster_consensus} offer quite a straightforward method to verify group consensusability of an MAS by comparing properties of the full underlying graph and its quotient graph. As mentioned in the Introduction, previous studies of group/cluster consensus problems such as \cite{Lu&Chen2010,Shang2016,Han&Chen2013,Han&Chen2015} rely on the condition of containing cluster spanning trees w.r.t. a clustering for $\cG$.
In the following, it is interesting to show in Proposition \ref{thm:equalnoofspanningtrees} that this condition is actually equivalent to that in Theorem \ref{thm:intracluster_consensus} after in-depth inspections on the relations between $\cG$ and its quotient graph $\GG$.

Subsequent presentation needs the following definitions \cite{Caughman&Veerman2006}.
For any node $i$ in $\GG$, a set $\RR(i)$ is a reachable set of $i$ if it contains $i$ and all nodes $j$ that can be reached starting from $i$ via a directed path in $\GG$.
A set $\RR_p$ is called a \emph{reach} if $\RR_p=\RR(i)$ for some $i$ and there is no $j$ such that $\RR(i)\subset\RR(j)$, and the node $i$ is called a root of this reach. Suppose $\RR_p,p=1,\ldots,m$, are the reaches that together cover all nodes of $\GG$. It is clear that if $\GG$ contains $m$ reaches, then its Laplacian $\cL_{\GG}$ is $m$-reducible.
For each reach $\RR_p$, the set $\VV_p=\RR_p\setminus\cup_{q\neq p}\RR_q$ is called the \emph{exclusive} part of $\RR_p$, and the set $\FF_p=\RR_p\setminus\VV_p$ denotes the \emph{common} part of $\RR_p$. Let $\FF=\cup_{p=1}^m \FF_p$ be the union of the common parts. Then, there exists a labeling of nodes of $\GG$ such that its Laplacian can be written into the following lower-triangular form \cite{Caughman&Veerman2006}
\begin{equation} \label{var:laplacian_reach1}
\cL_{\GG}=\left[ \begin{array}{llll}
V_{1}      &       &            &       \\
0          &\ddots &            &       \\
0          &0      & V_{m}      &       \\
F_{1}      &\cdots & F_{m}      &F      \\
\end{array} \right]
\end{equation}
where each $V_p, p=1,\ldots,m$ is a Laplacian matrix associated with $\VV_p$, $F$ is a square matrix associated with $\FF$, and $F_p$'s are matrices of compatible dimensions.

Now we can present the following lemma and Proposition \ref{thm:equalnoofspanningtrees} which will be used in the remaining parts of this paper.  Their proofs can be found in Appendix \ref{proof_of_lemma6}.
\begin{lemma}\label{thm:lemma6}
Suppose the Laplacian $\cL$ of $\cG$ satisfies Assumption \ref{assump:constant_row_sums}, and the associated $\cL_{\GG}$ takes the form \eqref{var:laplacian_reach1} for some $1\leq m< N$. Then the graph $\cG$ contains cluster spanning trees w.r.t. $\cC$ if and only if for each set of subgraphs $\{\cG_i|i\in\VV_p\}$, $p=1,\ldots,m$, the nodes therein can be spanned by a directed spanning tree in $\cG$. 
\end{lemma}

\begin{proposition} \label{thm:equalnoofspanningtrees}
Suppose the Laplacian $\cL$ of $\cG$ satisfies Assumption \ref{assump:constant_row_sums}. The spanning forest of $\cG$ and that of the quotient graph $\GG$ w.r.t. $\cC$ have equal minimum number of directed trees if and only if $\cG$ contains cluster spanning trees w.r.t. $\cC$.
\end{proposition}

A direct combination of Theorem \ref{thm:intracluster_consensus} and Proposition \ref{thm:equalnoofspanningtrees} leads to the following alternative of Theorem \ref{thm:intracluster_consensus}.
\begin{theorem}\label{thm:intracluster_consensus_corl}
Under Assumption \ref{assump:constant_row_sums}, the multi-agent system \eqref{eq:MAS_compact} can achieve group consensus w.r.t. $\cC$ for any initial state $x(0)\in\RR^{nL}$ if and only if $\cG$ contains cluster spanning trees w.r.t.  $\cC$.
\end{theorem}

\begin{remark}
In the literature, containing cluster spanning trees for a directed underlying graph is found to be a sufficient graph condition for achieving group consensus for single integrators \cite{Han&Chen2013,Han&Chen2015}. Its necessity is revealed only for special graphs, e.g., undirected graphs \cite{Lu&Chen2010}, balanced digraphs \cite{Shang2016}, and digraphs with the quotient graph $\GG$ being acyclic \cite{Liu&Wong_CDC2016}.
Theorem \ref{thm:intracluster_consensus_corl} consolidates this condition as a necessary and sufficient one for general nonnegative digraphs. 
In comparison to checking cluster spanning trees, the newly derived connectivity condition in Theorem \ref{thm:intracluster_consensus} is easier to check since the coupling details inside the clusters are not involved. Moreover, the condition of comparing the number of zero eigenvalues of the two Laplacians  is also a straightforward algebraic criterion.
\end{remark}

\section{Consensus States in Clusters} \label{sec:syn_states}
As is know, if the underlying topology $\cG$ contains a directed spanning tree and the inter-agent couplings are strong enough, the MAS \eqref{eq:MAS_compact} can achieve global consensus  with $x(t)\rightarrow(\ones_{L}\nu^T \otimes e^{At})x(0)$ where $\nu\in\RR^{L}$ is the left eigenvector of $\cL$ such that $\nu^T\cL=0$ and $\nu^T \ones_{L}=1$ \cite{Li&GRChen2010}. In this section, we are interested to see the consensus states in different clusters when the underlying graph of an MAS should be spanned by multiple trees together.
To this end, we assume the Laplacian $\cL_{\GG}$ is in the form of \eqref{var:laplacian_reach1} for some $1< m<N$. Then, the corresponding Laplacian $\cL$ of digraph $\cG$ can be written into the following form:
\begin{equation} \label{var:laplacianpartition_frobenius2}
\cL\!=\!\left[\begin{array}{llllll}
\bL_{1}\!    &       &            &           \\
           &\!\ddots\! &            &          \\
\zeros\!     &       & \!\bL_{m}\!    &           \\
\bL_{m+1,1}\!&\!\cdots\! &\! \bL_{m+1,m}\!&\! \cL_{\cF}
\end{array}
\right]
\end{equation}
where each $\bL_p$, $p=1,\ldots, m$, is the Laplacian associated with nodes in $\bar\cC_p=\cup_{i\in\VV_p}\cC_{i}$, and $ \cL_{\cF}$ is a square matrix associated with nodes in $\cF=\cup_{i\in\FF}\cC_{i}$.

\begin{lemma} \label{thm:invertible_submatrix}
If $\cG$ contains cluster spanning trees, then each $\bL_p$ for $p=1,\ldots,m$ contains exactly one zero eigenvalue, and the matrix $\cL_{\cF}$ is nonsingular with all eigenvalues having positive real parts.
\end{lemma}
\begin{IEEEproof}
The first half part follows immediately from Lemma \ref{thm:lemma6} and Lemma \ref{thm:m_reducible}. Since $\cL_{\GG}$ is $m$-reducible, $\cL$ is also $m$-reducible and has $m$ zero eigenvalues totally according to Proposition \ref{thm:equalnoofspanningtrees} and Lemma \ref{thm:m_reducible}. Therefore, $\cL_{\cF}$ must be nonsingular.
\end{IEEEproof}

\subsection{Consensus States with Strong Coupling Strengths}
Denote $\cR=\cup_{p=1}^{m}\bar\cC_p$, and let $\cL_{\cR}=blkdg\{\bL_1,\ldots,\bL_m\}$ and $\cL_{\cF\cR}=[\bL_{m+1,1},\ldots, \bL_{m+1,m}]$. Then \eqref{var:laplacianpartition_frobenius2} can be rewritten as
\begin{equation}\label{eq:CS_9}
\cL=\left[\begin{array}{ll}
      \cL_{\cR}&\zeros\\ \cL_{\cF\cR}& \cL_{\cF}
    \end{array}\right].
\end{equation}
Similarly, the state vector $x(t)$ is also represented as follows
 \begin{equation}\label{eq:CS_10}
 x(t)=[x^T_{\cR}(t),x^T_{\cF}(t)]^T.
 \end{equation}
Then, we can derive the final states in each cluster when the overall coupling strength $\delta$ is large enough.
\begin{theorem} \label{thm:intra-cluster-states}
Under Assumption \ref{assump:constant_row_sums}, if the underlying graph $\cG$ of the MAS \eqref{eq:MAS_compact} contains cluster spanning trees w.r.t. $\cC$, by selecting $K=B^TP$ and
\begin{equation} \label{eq:CS_14}
\delta\ge\dfrac{1}{2\underline{\lambda}},\;\; \underline{\lambda}:=\min \{Re\lambda_{l}(\cL):\lambda_{l}(\cL)\ne 0,\forall l\in\cI\},
\end{equation}
the state $x(t)$ will asymptotically approach the following
\begin{subequations}\label{eq:CS_13}
\begin{numcases}{}
(\ones_{L}\nu^T \otimes e^{At})x(0), \textrm{if $\cL$ is irreducible or $1$-reducible}\label{eq:CS_13a}\\
\begin{bmatrix}
\Xi\otimes e^{At} \\
-\cL_{\cF}^{-1}\cL_{\cF\cR}\Xi\otimes e^{At}
\end{bmatrix} x(0), \;\textrm{if $\cL$ has the form \eqref{var:laplacianpartition_frobenius2}}\label{eq:CS_13b}
\end{numcases}
\end{subequations}
as $t\rightarrow \infty$, where $\Xi=blkdg\{\mu_{1}\nu^T_1,\ldots,\mu_{m}\nu^T_m\}$ with $\mu_p$ and $\nu_p$ satisfying $\nu_p^T\mu_p=1$ being the right and left eigenvector of $\bL_p$ associated with the zero eigenvalue, respectively. In the special case that $A=0,B=1,K=1$, i.e., single integrators, $x(t)$  will reach \eqref{eq:CS_13} with $A=0$ for any $\delta>0$.
\end{theorem}

\begin{IEEEproof}
We only need to prove the case that $\cL$ is reducible and takes the form of \eqref{var:laplacianpartition_frobenius2} for some $m>1$.
For each $p=1,\ldots, m$, denote $\bar x_{p}$ as the stacked states of all agents in the union of clusters $\bar\cC_p=\cup_{i\in\VV_p}\cC_{i}$.
It follows from \eqref{eq:MAS_compact} and \eqref{var:laplacianpartition_frobenius2} that
\begin{align} \label{eq:CS_15c}
\dot {\bar x}_{p}(t)=(I_{n_p}\otimes A-\delta\bL_{p}\otimes BK) {\bar x}_{p}(t),\;  p=1,\ldots, m
\end{align}
where $n_p$ is the dimension of $\bL_{p}$.
If  $\cG$ contains cluster spanning trees w.r.t. $\cC$, by Lemma \ref{thm:invertible_submatrix}, $\bL_p$ has a unique zero eigenvalue $\lambda_{1}(\bL_p)=0$, and other eigenvalues satisfy
$\min_{l_p\neq 1} Re\lambda_{l_p}(\bL_p)\ge \underline{\lambda}$. By  \eqref{eq:CS_14}, there holds $\delta\ge1/\min_{l_p\neq 1} 2 Re\lambda_{l_p}(\bL_p)$. Going through a similar algebra
as in \eqref{eq:CS_7} yields that each $A -\lambda_{l_p}(\bL_p)BK, l_p\ne 1$ is Hurwitz. Hence, one can use methods in Theorem 2 of \cite{Li&GRChen2010} to solve \eqref{eq:CS_15c} and get that for $p=1,\ldots, m$,
\begin{equation}\label{eq:CS_16a}
\bar x_{p}(t)\rightarrow(\mu_{p}\nu^T_p\otimes e^{At})\bar x_{p}(0),\; \textrm{as}\; t\rightarrow \infty.
\end{equation}
It then follows from the definition of $\cR$ and \eqref{eq:CS_10} that
\begin{equation}\label{eq:CS_16}
x_{\cR}(t)\rightarrow(\Xi\otimes e^{At})x_{\cR}(0),\; \textrm{as}\; t\rightarrow \infty.
\end{equation}

To derive the state of $x_{\cF}(t)$ when $t\rightarrow \infty$, we define the following two variables following \eqref{eq:CS_9} and \eqref{eq:CS_10}:
\begin{align} \label{eq:CS_20}
\begin{bmatrix} \zeta\\ \xi\end{bmatrix}=(\cL\otimes I_n) x =\left[\begin{array}{ll}
      \cL_{\cR}&\zeros\\ \cL_{\cF\cR}& \cL_{\cF}
    \end{array}\right]\begin{bmatrix} x_{\cR}\\ x_{\cF}\end{bmatrix}.
\end{align}

By \eqref{eq:CS_16} and using the fact $\cL_{\cR}\Xi=0$, one has that
\begin{align} \label{eq:CS_23}
\zeta(t)&=(\cL_{\cR}\otimes I_n)x_{\cR} \notag\\
&\rightarrow(\cL_{\cR}\otimes I_n)(\Xi\otimes e^{At})x_{\cR}(0)
=(\cL_{\cR}\Xi\otimes e^{At})x_{\cR}(0)\notag\\
&=0,\;\textrm{as}\; t\rightarrow \infty.
\end{align}

Using \eqref{eq:MAS_compact} and \eqref{eq:CS_20}, we have that
\begin{align}\label{eq:CS_17}
\begin{bmatrix} \dot \zeta\\ \dot \xi\end{bmatrix}
&=(\cL\otimes I_n) \dot x =(\cL\otimes I_n) (I_L\otimes A-\delta\cL\otimes BK) x(t)\notag\\
&=(I_L\otimes A-\delta\cL\otimes BK) \begin{bmatrix} \zeta\\ \xi\end{bmatrix}.
\end{align}
It follows from \eqref{eq:CS_9} and \eqref{eq:CS_10} that
\begin{align}
 \label{eq:CS_17b}
\dot \xi&=(I\otimes A-\delta\cL_{\cF}\otimes BK) \xi-(\delta\cL_{\cF\cR}\otimes BK)\zeta.
\end{align}
By Lemma \ref{thm:invertible_submatrix}, there holds $Re\lambda_{l}(\cL_{\cF})>0, \forall l$. It follows that $\underline{\lambda}\le \min_{l} Re\lambda_{l}(\cL_{\cF})$, which combining \eqref{eq:CS_14} implies that $\delta\ge 1/\min_{l} 2 Re\lambda_{l}(\cL_{\cF})$. Then, through a similar algebra as in \eqref{eq:CS_7}, one can get that $A-\delta\lambda_{l}(\cL_{\cF})BK=A-\delta\lambda_{l}(\cL_{\cF})BB^TP$ is Hurwitz for each $\lambda_{l}(\cL_{\cF})\in\sigma(\cL_{\cF})$. That is, $I\otimes A-\delta\cL_{\cF}\otimes BK$ is Hurwitz.
Next, solving \eqref{eq:CS_17b} with \eqref{eq:CS_23}, one can obtain that $\xi(t)$ approaches zero asymptotically. Since $\xi=\cL_{\cF\cR}x_{\cR}+\cL_{\cF}x_{\cF}$ by \eqref{eq:CS_20}, it then follows from \eqref{eq:CS_16} that
\begin{align}
x_{\cF}(t)&\rightarrow-(\cL_{\cF}^{-1}\cL_{\cF\cR}\otimes I_n)x_{\cR}(t)\notag\\
\label{eq:CS_22}
&\rightarrow-(\cL_{\cF}^{-1}\cL_{\cF\cR}\Xi\otimes e^{At})x_{\cR}(0), \; \textrm{as}\;  t\rightarrow \infty.
\end{align}
Combining \eqref{eq:CS_16} and \eqref{eq:CS_22} yields the state of $x(t)$ in \eqref{eq:CS_13b} when $\cL$ takes the form \eqref{var:laplacianpartition_frobenius2}.
For the case that $A=0,B=1,K=1$, it is straightforward to check that the above proofs are valid for any $\delta>0$.
This completes the proof.
\end{IEEEproof}

\begin{remark}
As seen from \eqref{eq:CS_16a}, the clusters $\cC_i$'s with ${i\in\VV_p}$ eventually achieve a common consensus state;
for clusters in $\cF$, note from \eqref{eq:CS_22} that their states $x_{\cF}(t)$ eventually enter into the convex hull of $x_{\cR}(t)$. 
To see this, by $[\cL_{\cF\cR}\;\cL_{\cF}]\ones_{L}=\cL_{\cF\cR}\ones_{|\cR|}+\cL_{\cF}\ones_{|\cF|} =\zeros$, one has $-\cL_{\cF}^{-1}\cL_{\cF\cR}\ones_{|\cR|}=\ones_{|\cF|}$ where $-\cL_{\cF\cR}$ is a nonnegative matrix and $\cL_{\cF}^{-1}$ is also a nonnegative matrix since $\cL_{\cF}$ is a nonsingular $M$-matrix \cite{Plemmons1977M_matrix}. Hence, $-\cL_{\cF}^{-1}\cL_{\cF\cR}$ is row stochastic. This group consensus pattern is consistent with that of MASs described by single integrators (i.e., $A=0,B=1$) under unweighted digraphs\cite{Monaco2019}. However, for a nontrivial system matrix $A$, the occurrence of the pattern \eqref{eq:CS_13} also relies on a large enough coupling strength $\delta$, e.g., that satisfying \eqref{eq:CS_14}. A counterexample is when $\delta$ is decreased such that $\frac{1}{\min 2 Re\lambda(\hat \cL)}\le \delta< \frac{1}{2\underline{\lambda}}$, the generic MAS \eqref{eq:MAS_compact} can still achieve group consensus by Theorem \ref{thm:intracluster_consensus} but there is no guarantee for  $x_{\cF}(t)$ to enter the convex hull of $x_{\cR}(t)$ as illustrated by a simulation example in the next subsection. This is because the achieved multiple consensus states don't converge due to unstable modes of $A$, and inter-cluster coupling strengths are not strong enough to further synchronize them.

Further notice that \eqref{eq:CS_13} contains the minimum number of distinct consensus states that can persist under a directed graph, i.e., the consensus states of different clusters won't be merged further by increasing the overall coupling strength $\delta$. Particularly if the underlying graph can be spanned by one directed tree, all systems' states will merge into one consensus state when $\delta$ is large enough, i.e., reach global consensus, as seen from \eqref{eq:CS_13a}.  This phenomenon also appears in connected undirected networks \cite{OClery2013,Schaub2016,Gambuzza&Frasca2019}. Hence, it is trivial to see that Theorem \ref{thm:intra-cluster-states} includes global consensus as a special case.
\end{remark}

\subsection{Simulation Example}
\begin{figure}[th]
  \centering
  \includegraphics[width=7cm]{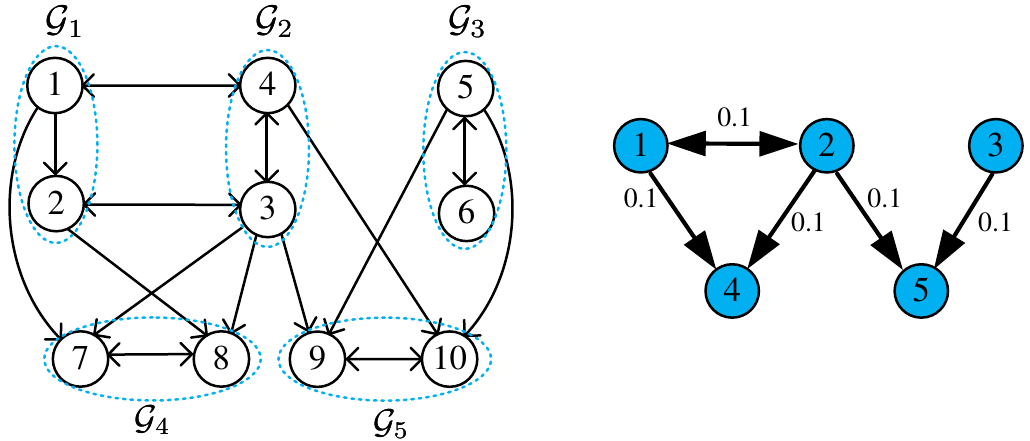}
  \caption{The graph $\cG$ (on the left) and its quotient graph $\GG$ (on the right). $\cG$ consists of five clusters of nodes with all intra-cluster edge weights equal to $1$ and all inter-cluster edge weights equal to $0.1$.}
  \label{fig:topology_simu}
\end{figure}

To illustrate the consensus states, we present a simulation example for an MAS consisting of $10$ agents that belong to $5$ clusters $\cC_1=\{1,2\}$, $\cC_2=\{3,4\}$, $\cC_3=\{5,6\}$, $\cC_4=\{7,8\}$, $\cC_5=\{9,10\}$. 
The underlying graph $\cG$ is given in Fig. \ref{fig:topology_simu}, which contains cluster spanning trees w.r.t. the clustering $\cC=\{\cC_1,\cC_2,\cC_3,\cC_4,\cC_5\}$, and its Laplacian $\cL$ satisfies Assumption \ref{assump:constant_row_sums}.
The dynamics of the agents are described by harmonic oscillators, that is, for $i=1,\ldots,10$,
\begin{subequations}\label{eq:CS_24}
\begin{numcases}{}
\dot x_{1l(t)}= x_{2l}(t)\\
\dot x_{2l(t)}= -x_{1l}(t)+u_l(t).
\end{numcases}
\end{subequations}
Selecting $Q=I$, and solving the algebraic Riccati equation \eqref{eq:CS_8}, we obtain the controller gain $K=B^TP=[0.4142,    1.3522]$.

\begin{figure}[th]
  \centering
  \includegraphics[width=8.5cm]{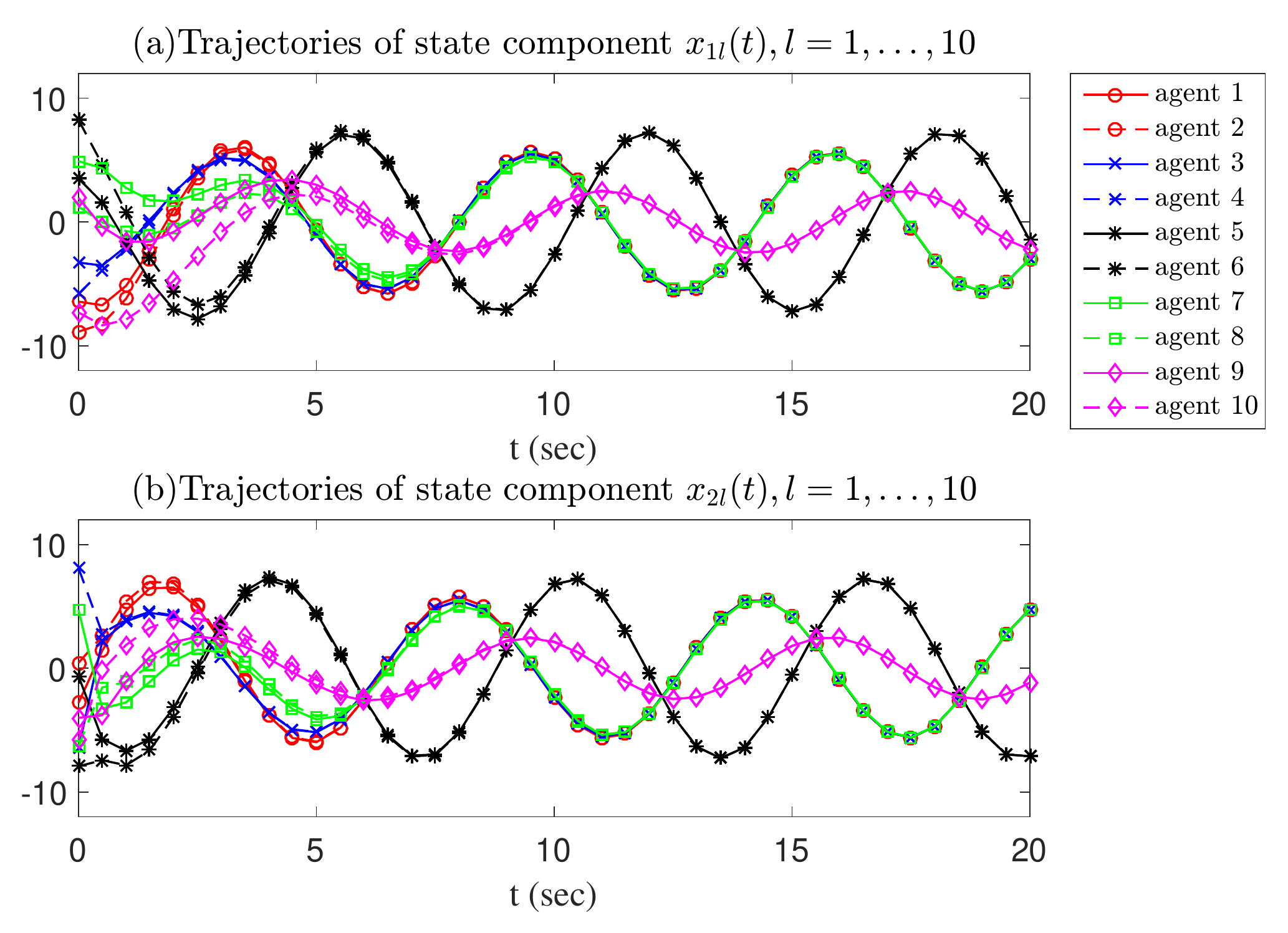}
  \caption{The 10 agents achieve group consensus and form 3 distinct consensus states when $\delta=1/2\underline{\lambda}$. }
  \label{fig:simu_3clusters}
\end{figure}

\begin{figure}[ht]
  \centering
  \includegraphics[width=8.5cm]{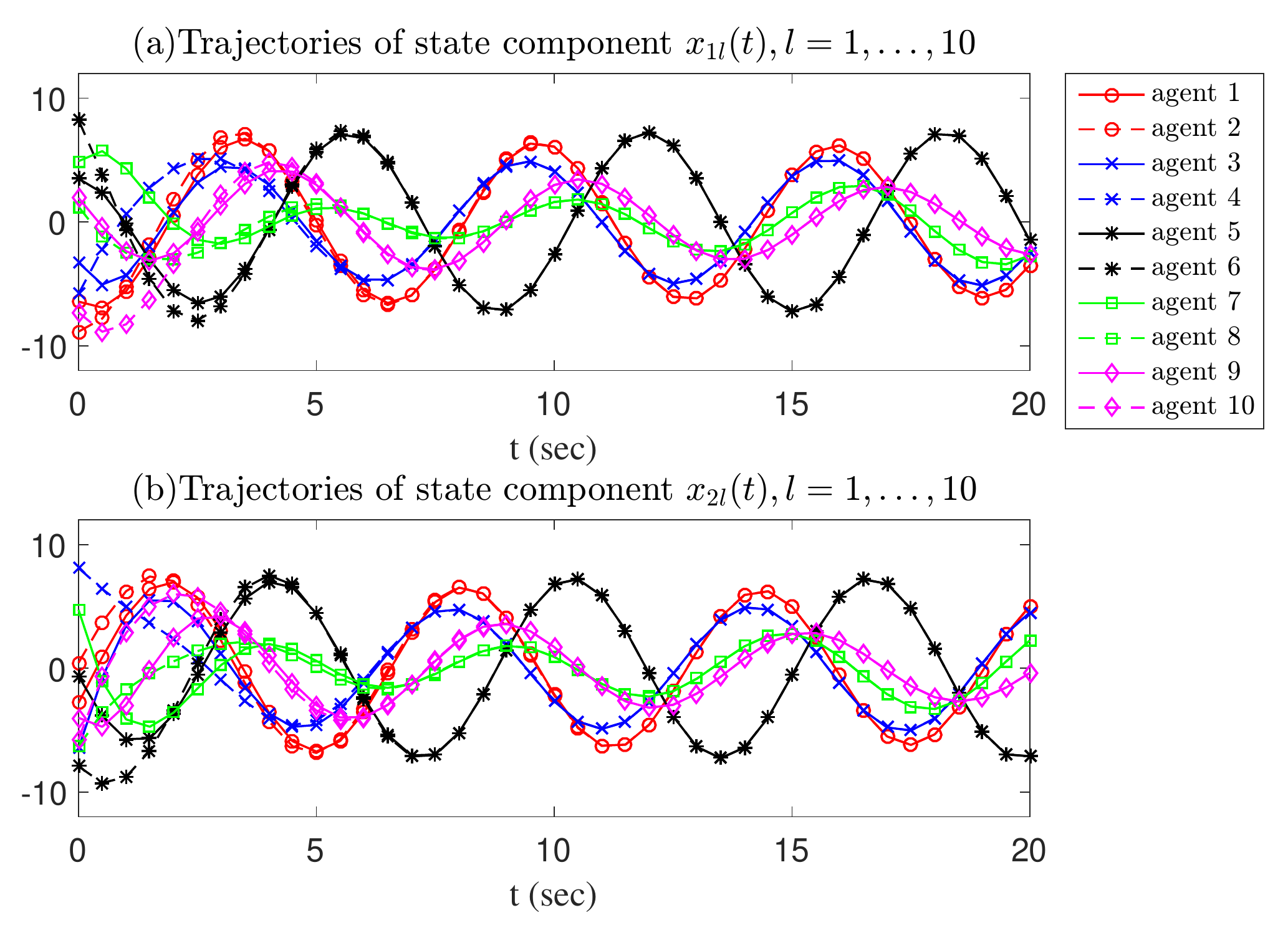}
  \caption{The 10 agents achieve group consensus and form 5 distinct consensus states when $\delta=1/\min 2 Re\lambda(\hat \cL)<1/2\underline{\lambda}$.}
  \label{fig:simu_5clusters}
\end{figure}

It is computed that $\min Re\lambda(\hat \cL)=1.09>0$ and $\underline{\lambda}=0.2$. Hence, we first set $\delta=1/(2\underline{\lambda})=2.5$ according to \eqref{eq:CS_14} in Theorem \ref{thm:intra-cluster-states}.
With randomly generated initial states, the simulated trajectories of the $10$ agents are shown in Fig. \ref{fig:simu_3clusters}, in which the states of agents form three groups in such a way that clusters $\cC_1$, $\cC_2$, and $\cC_4$ merge into one state, while $\cC_3$ and $\cC_5$ each achieves a distinct consensus state. Note also that the consensus states of $\cC_5$ lie in between the states of clusters $\cC_1$ and $\cC_3$ when $t$ is large enough.
Next, we use a smaller value for $\delta$ by setting $\delta=1/\min 2 Re\lambda(\hat \cL)=0.4587$ according to \eqref{eq:CS_14a}. Simulation results in Fig. \ref{fig:simu_5clusters} show that five groups of distinct states are formed eventually complying with the partition $\cC$, i.e., the success of achieving group consensus, but no evident relations can be observed for the consensus states in different clusters.


\section{Conclusions} \label{sec:conclusion}
We have investigated the group consensus problem for generic linear multi-agent systems under nonnegative directed graphs. A necessary and sufficient condition in terms of the topologies of the underlying digraph and its quotient graph is presented. This condition is shown to be equivalent to an existing condition commonly used for undirected graph. Thus, a unified understanding of the graph topologies for ensuring group consensus is established. The consensus states in different clusters are also presented explicitly when the overall coupling of the underlying graph is strong enough, thanks to the identical individual linear system models. It is shown by both theoretical analysis and simulation examples that the coupling strengths for achieving group consensus and for realizing the minimum number of distinct consensus states could be different.  Generally, the final group consensus pattern is an outcome of the interplay of the agents' individual dynamics, the underlying graph topology and the overall coupling strength $\delta$, and thus is hard if not impossible to allow an accurate or explicit specification. For example, the group consensus pattern in \eqref{eq:CS_13} or \cite{Monaco2019} for identical linear systems may not appear at all in MASs with nonidentical linear dynamics in different clusters as observed in the simulation results of our previous work \cite{Liu&Wong_CDC2016}. Although existing works such as \cite{Pecora2014Nature,Sorrentino&Pecora2016,Schaub2016,Klickstein&Pecora2019} have exemplified the occurrence of different patterns in coupled nonlinear oscillators via simulations, a specific characterization remains a challenging task that needs further efforts.
Another interesting future work is to cope with noncooperative agents with generic linear dynamics or nonlinear dynamics by using resilient cluster censoring strategies as proposed in a recent paper \cite{Shang2021} for scalar systems, in which the inter-cluster common influence condition is released.

\appendices
\section{}\label{proof_of_lemma6}
In order to prove Lemma \ref{thm:lemma6} and Proposition \ref{thm:equalnoofspanningtrees}, we need the following preliminary results.
\begin{lemma} \label{thm:lemma3}
 If $\cG$ contains a directed spanning tree (is strongly connected), then $\GG$ also contains one (is strongly connected).
\end{lemma}

\begin{lemma}\label{thm:lemma5}
Under Assumption \ref{assump:constant_row_sums}, if $\GG$ has a directed spanning tree and its root node is associated with a subgraph $\cG_i$ of $\cG$ that has a directed spanning tree, then $\cG$ has a directed spanning tree.
\end{lemma}
\begin{IEEEproof}
For the spanning tree of $\GG$, suppose without loss of generality that its root node is associated with subgraph $\cG_1$ in $\cG$. Note that each directed link of the spanning tree of $\GG$ is associated with inter-cluster links in $\cG$ pointing from one subgraph to another. Hence, every subgraph $\cG_i$, $i\neq 1$ is pointed by inter-cluster links originating from some other subgraph. Moreover, every node in each $\cG_i$, $i\neq 1$ is pointed by at least one inter-cluster link due to Assumption \ref{assump:constant_row_sums}. Hence, there exists a path from the subgraph $\cG_1$ to all nodes outside $\cG_1$ via inter-cluster links that are associated with the links of the spanning tree of $\GG$. Note that this path can be an extension of a path in the spanning tree of $\cG_1$. It follows that $\cG$ contains a directed spanning tree with its root being the root of the spanning tree of $\cG_1$.
\end{IEEEproof}

\begin{lemma}\label{thm:lemma4}
Under Assumption \ref{assump:constant_row_sums}, if $\GG$ is strongly connected and there exists a subgraph $\cG_i$ of $\cG$ whose nodes can be spanned by a directed tree in $\cG$, then $\cG$ contains a directed spanning tree. 
\end{lemma}
\begin{IEEEproof}
The strong connectivity of $\GG$ implies that every node in each subgraph $\cG_j$, $j\in\{1,2,\ldots,N\}$ of $\cG$ is pointed by inter-cluster links originating from at least one other subgraph $\cG_{j'}$, $j'\neq j$. Using similar arguments as those in the proof of Lemma \ref{thm:lemma5}, one sees that the directed tree that spans $\cG_i$ can be expanded to reach all nodes in $\cG$ through inter-cluster links.
\end{IEEEproof}

\subsection{Proof of Lemma \ref{thm:lemma6}}
\begin{IEEEproof}
The necessity part follows from Lemma \ref{thm:lemma5} by using the definitions of cluster spanning trees and the set $\VV_p$.
For the sufficiency part, denote by $\cT_p$ for $p=1,\ldots,m$ the directed spanning tree that contains all nodes in $\{\cG_i|i\in\VV_p\}$. Note that $\VV_p$ shares the same root node with the reach $\RR_p$. The if part of this lemma implies that the subgraph $\cG_i$ associated with this root node of $\RR_p$ is spanned by a component of the directed tree $\cT_p$. It follows from Lemma \ref{thm:lemma5} that any cluster of nodes $\cC_i$ with $i\in\RR_p$ can be spanned by a directed tree (which contains $\cT_p$). The proof is completed when noting that the reaches $\RR_1,\ldots,\RR_m$ contain the labels of all clusters.
\end{IEEEproof}
\subsection{Proof of Proposition \ref{thm:equalnoofspanningtrees}}
\begin{IEEEproof} 
Suppose the minimum number of directed trees which together span $\GG$ is $m$. Then the proof of this proposition is converted to showing the equivalence of the following two statements:
\begin{itemize}
   \item[(a)] $\cG$ contains cluster spanning trees w.r.t. $\cC$.
   \item[(b)] the minimum number of directed trees which together span $\cG$ is $m$.
\end{itemize}
For $m=1$, this equivalence has been established by combing Lemma \ref{thm:lemma3} to Lemma \ref{thm:lemma4}. For $1<m<N$, considering the subset of subgraphs in $\{\cG_i|i\in\cup_{p=1}^{m}\VV_p\}$, one needs at least $m$ directed trees in order to span all of the nodes therein (at least one directed spanning tree for each set of subgraphs $\{\cG_i|i\in\VV_p\}$).

(a) $\Rightarrow$ (b): By the necessity part of Lemma \ref{thm:lemma6} and its proof, 
$m$ is a feasible number of directed spanning trees that together span $\cG$. Hence, statement (b) holds.

(b) $\Rightarrow$ (a): If (a) does not hold, then according to Lemma \ref{thm:lemma6} there exists a $p^*\in\{1,\ldots,m\}$ such that the nodes of $\{\cG_i|i\in\VV_{p^*}\}$ cannot be spanned by any single tree. It follows that more than $m$ directed trees are needed in order to span all of the nodes in $\{\cG_i|i\in\cup_{p=1}^{m}\VV_p\}$, i.e., the negation of statement (b) is true. Hence, (b) $\Rightarrow$ (a) holds.
\end{IEEEproof}
\addtolength{\textheight}{0cm}   



\end{document}